\documentclass[preprint,12pt,authoryear]{elsarticle}

\usepackage{amssymb}
\usepackage{lipsum}

\journal{Astronomy $\&$ Computing}

\begin{document}

\begin{frontmatter}



\title{Automatic detection of solar flares observed at 45 GHz by the POEMAS telescope}


\author[first]{Vanessa Lessa}
\author[first]{Adriana Valio}
\affiliation[first]{organization={Center for Radio Astronomy and Astrophysics Mackenzie, Mackenzie Presbyterian University},
            addressline={Rua da Consolacao, 930}, 
            city={São Paulo},
            postcode={01302-907}, 
            state={SP},
            country={Brazil}}

\begin{abstract}
Every 11 years, the Sun goes through periods of activity, with the occurrence of many solar flares 
and mass ejections, both energetic phenomena of magnetic origin. Due to its effects on Earth, the study of solar activity is of paramount importance. POEMAS (Polarization of Millimeter Emission of Solar Activity) is a system of two telescopes, installed at CASLEO (El Leoncito Astronomical Complex) in Argentina, which monitors the Sun at two millimeter wavelengths (corresponding frequencies of 45 and 90 GHz). The objective of this work is to automatically detect  solar flares observed by the polarimeter. First it is necessary to eliminate the background noise, caused mainly by instrumental problems, from the light curves of millimeter solar emission.
The methodology used to exclude the noise proposed in this work is to use the tendency of time series. 
The subtraction of this model from the light curves provides the input to automate the detection of solar flares using artificial intelligence techniques. A Neural Network was trained to recognize patterns and analyze a dataset in order to identify solar flares. 
Previously, a total of 30 flares had been visually identified  and analyzed in the POEMAS database  between 2011/11/22 and 2013/12/10. The methodology presented here confirmed 87\% of these events, moreover  the neural network was able to identify at least 9 new events.
As the neural network was trained to detect impulsive events (lasting less than 5 min), long duration bursts were not automatically detected, nor were they detected visually due to the background noise of the telescope. Visual inspection of the POEMAS data, when comparing with microwave data from the RSTN, allowed the identification of an additional 10 long-duration solar flares at 45 GHz.
We discuss some problems encountered and possible solutions for future work.
\end{abstract}




\end{frontmatter}
\tableofcontents


\section{Introduction}
The Sun is an active star with a magnetic cycle of about 11 years \citep{hathaway2013}. In periods of maximum activity, an increase in the frequency of solar flares and coronal mass ejections can be observed. Both particles and magnetic fields thrown into interplanetary space by coronal mass from the Sun may impact the Earth. Geomagnetic storms, disruption of telecommunications signals, GPS malfunctions, and blackouts are some of the disruptions affecting Earth.

Over the years, studies on solar activity and the Sun's behavior have been carried out trying to mitigate these  effects on Earth \citep{pulkkinen2007}. For example, studies involving active regions, magnetic fields, solar flares, coronal mass ejections, and others are all relevant. Since the emission from solar activity is produced at all wavelengths of the eletromagnetic spectrum,  observations  at different frequencies is crucial to understand solar phenomena and the mechanisms involved \citep{dulk1985}.

In 1859, English astronomers Richard C. Carrington and Richard Hodgson identified the first solar flare \citep{tsurutani2003}. The explosion was quite intense, and a flash located in a small region was detected in images of the Sun's visible light. Just 17 hours later, a coronal mass ejection hit the Earth, causing one of the largest magnetic storms ever recorded. If a similar storm reaches our planet today, it would cause severe communication and electrical energy problems, among others \citep{phillips2014}. 

Solar phenomena are usually associated with active regions of the solar atmosphere. When a solar flare occurs, a large amount of energy is released ($10^{28} - 10^{32}$ erg), this energy is used in accelerating particles and heating the plasma, which generate radiation across the entire electromagnetic spectrum (from X rays to radio waves) \citep{mann2009}.

Observations of solar activity, both from ground and space  observatories, have generated a large amount of data. Thus it is necessary to apply artificial intelligence techniques to analyze the data in search of the sudden increases in the emission caused by solar flares. Here we have used a neural network to find patterns and identify solar flares automatically.

This work involves the automatic detection of solar flares in the data of the Polarization Emission of Millimeter Activity at the Sun (POEMAS) telescope \citep{valio2013}. POEMAS is a polarimeter that observed the Sun daily from December 2011 to December 2013 at the rarely explored frequencies of 45 and 90 GHz. 

The paper is organized as follows. In Section~\ref{sec:poemas}, we describe the data, and in Section~\ref{sec:NN}, the Neural Network methodology. In Section~\ref{sec:results}, the results of the Neural Network experiments are detailed. Finally, we conclude, in Section~\ref{sec:conclusion}, and anticipate future research.

\section{POEMAS Telescope}\label{sec:poemas}
POEMAS is a system of two telescopes, installed at the CASLEO Observatory (El Leoncito Astronomical Complex), in Argentina. 
The POEMAS telescope provides solar left and right circular polarization measurements at two-millimeter wavelengths (45 and 90 GHz) with a temporal resolution of 10ms. It operated continually for two years (Dec 2011 - Dec 2013) observing the full disk of the Sun, and detected several flares. The data collected from the Sun every day were written to binary files, converted using the Python programming language, and finally written to FITS files. 

\subsection{Data Acquisition}
The antenna temperature data at both left and right circular polarized emission at 45 and 90 GHz are recorded in the daily binary files (TRK extension). Also the azimuth and elevation angles of the Sun are  recorded in the file. For the analyzes, we used the light curve resulting from the sum of the two, right (RCP) and  left (LCP), circular polarizations of the antenna temperature at 45 GHz. Using the Python programming language version 3.6, the TRK files were converted to FITS files using the following procedures:
\begin{itemize}
    \item FITS level 0 - conversion of data from the TRK to FITS file, using the 10ms configuration of the original file.
\item FITS level 1 - integration of the temporal resolution of the FITS file level 0 from 1 ms to 1s using the median of the data within the 1s interval.
\item FITS level 2 - the merger of all FITS files level 1 of the same day into a single new file
\item FITS level 3 - application of the time series (Trend) for all FITS file level 2
\end{itemize}
The first day of POEMAS observation used in this work was 12/01/2011, while the last day used was 12/10/2013. In this period, we did not have data for 51 days, which resulted in  a total of 690 days for analysis.

After converting the POEMAS binary files to Flexible Image Transport System (FITS) files, the data is integrated into  the database. 
Finally, the automatic detection of solar flares is applied using Artificial Intelligence techniques, especially pattern recognition and machine learning (Deep Learning).

\subsection{Data calibration}
\begin{figure}
\centering
\includegraphics[width=.6\textwidth]{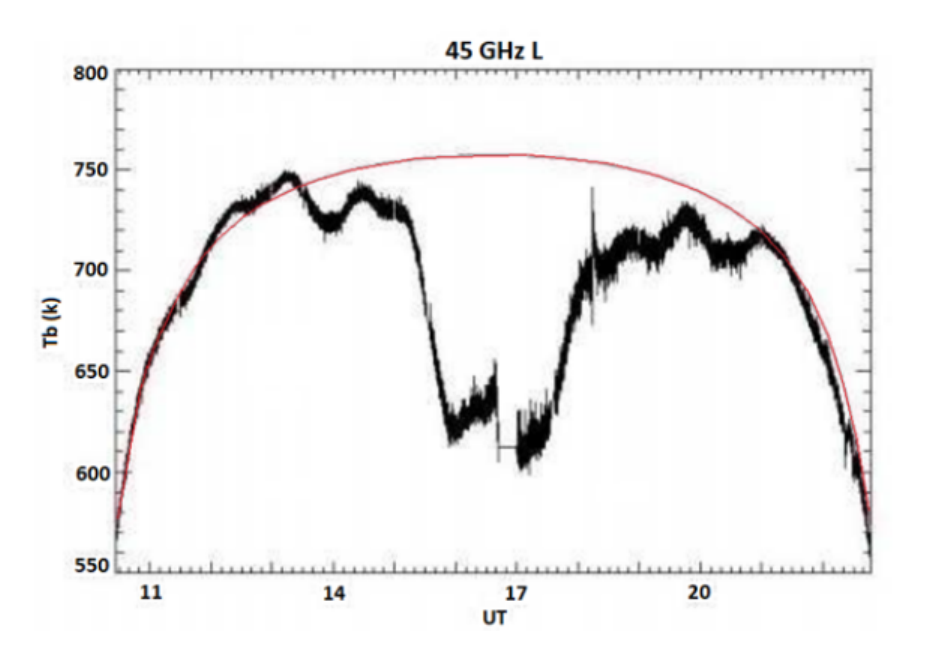}
\caption{Full-day observation of solar energy flux at 45 GHz, left circular polarization (black curve) and attenuated flux (red curve).}
\label{fig:Fig_Cap2_Obs45} 
\end{figure}

Unfortunately, there is a misalignment of the telescope support structure due to mechanical problems, causing the signal to abruptly decrease during local noon. This decrease of the antenna temperature is clearly seen in Figure~\ref{fig:Fig_Cap2_Obs45}, especially between 16 and 17 UT. The red curve on the same plot depicts the expected light curve profile of the observations.
Due to variations in the solar emission caused by this telescope misalignment, it is difficult to identify any increase caused by a solar event, except for the most intense ones, which are usually rare. 

\begin{figure}
\centering
\includegraphics[width=1\textwidth]{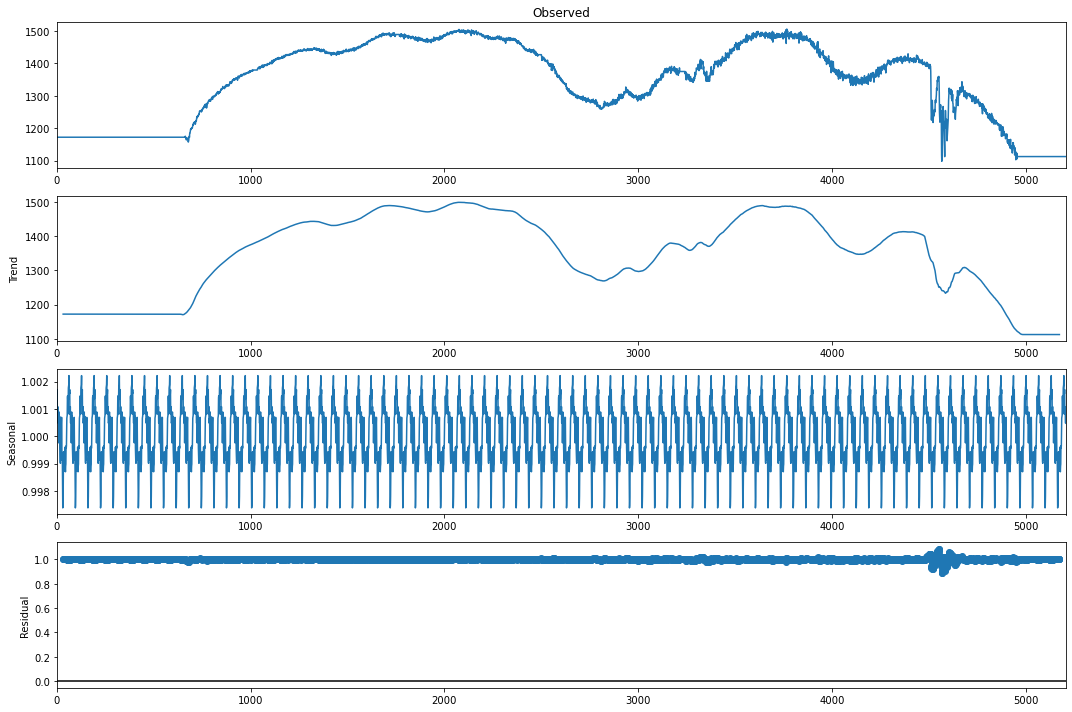}
\caption{Decomposition of the time series of 2012/01/27}
\label{fig:Fig_Cap5_Decomposicao} 
\end{figure}

Therefore, to minimize the daily variation of the telescope measurements at 45 GHz, a time series subtraction of the signal was performed. This time series is the component trend of the antenna temperature of the same polarization on the day under consideration.
In Figure~\ref{fig:Fig_Cap5_Decomposicao}, we can see the 3 components of the time series for the observations on 01/27/2012. In the top panel of Figure~\ref{fig:Fig_Cap5_Decomposicao}, the light curve observed by POEMAS, that is the result of the sum of the RCP and LCP polarizations, is shown. The second plot from the top is the  signal trend for different growth and decrease patterns. In the third plot, the seasonality is presented; in this case, we consider an interval of 50 points to analyze the behavior for every period of 1s. In the last panel, the residuals after subtraction of the effects of seasonality and trend from the data are presented. The residual fluctuations are attributed to random components.

\begin{figure}[h]
\centering
\includegraphics[width=1\textwidth]{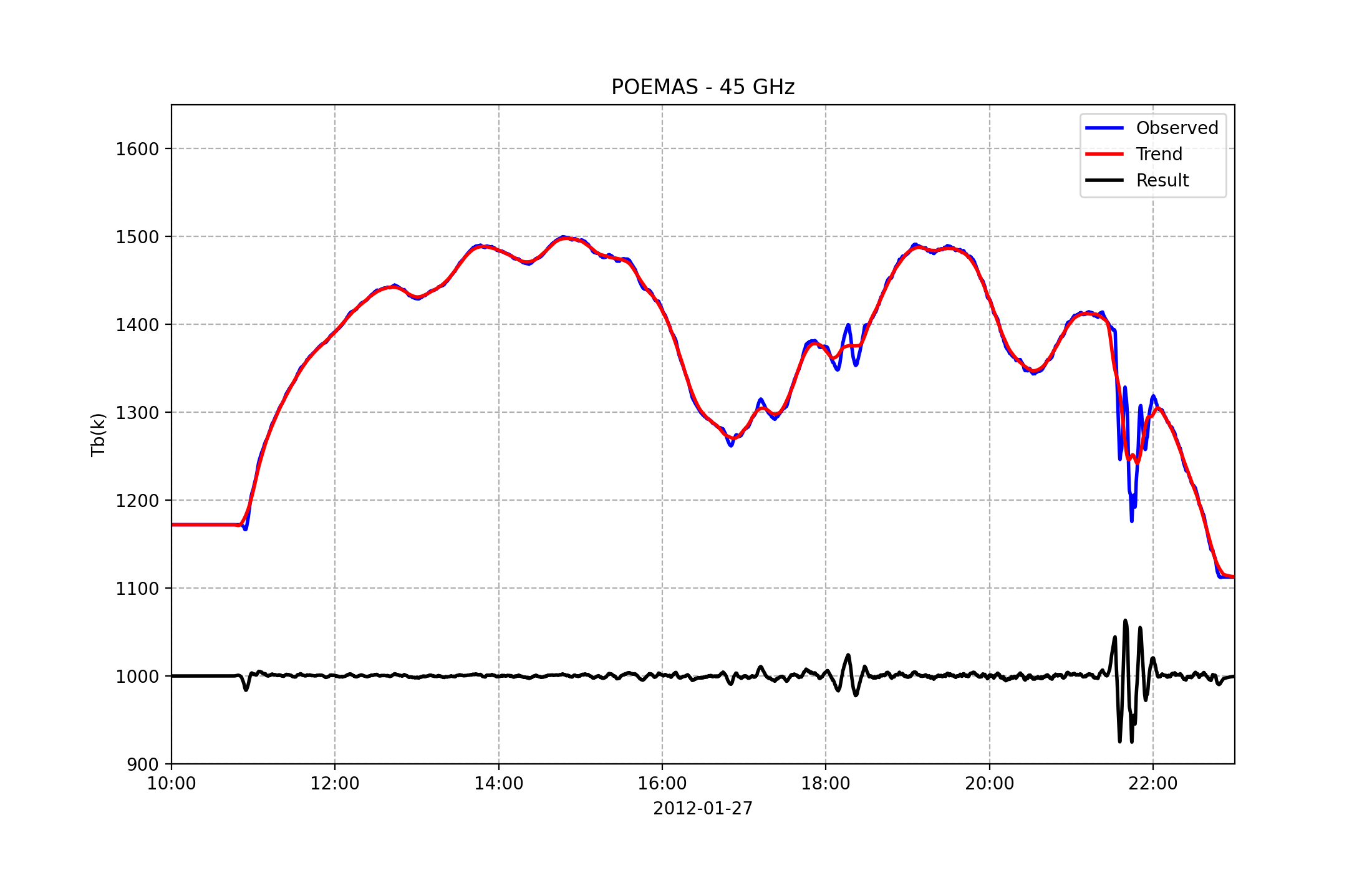}
\caption[Analysis of the emission observed by POEMAS at 45 GHz all day 2012/01/27]{Analysis of the emission observed by POEMAS at 45 GHz all day 2012/01/27} 
\label{fig:Fig_Cap5_Analise_ST} 
\end{figure}

In Figure~\ref{fig:Fig_Cap5_Analise_ST},  the original observed data (blue curve), the trend of the time series (red curve), and the result after subtracting the trend from the observed signal  (black curve) are shown. A value of 1000 was added to this subtraction residual to place it on the same scale as the original signal. On this day, a solar flare occurred at 18:15 UT, and the impulsive peak can be observed in the original curve (blue). On the same day, between 21:00 UT and 22:00 UT, there was a drop in signal due to interference in the Earth's atmosphere, probably due to clouds in front of the Sun. These drops in signal due to clouds are significant noise in the signal, and we have disregarded them from the data analysis.

\begin{figure}
\centering
\includegraphics[width=.9\textwidth]{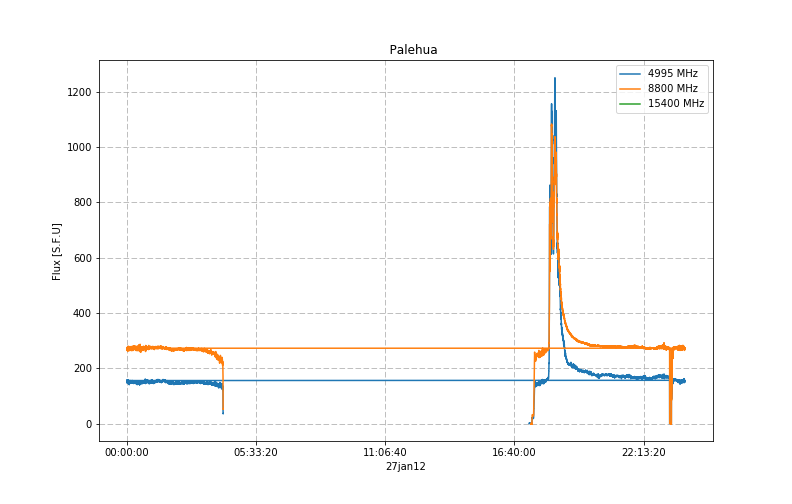}
\includegraphics[width=.9\textwidth]{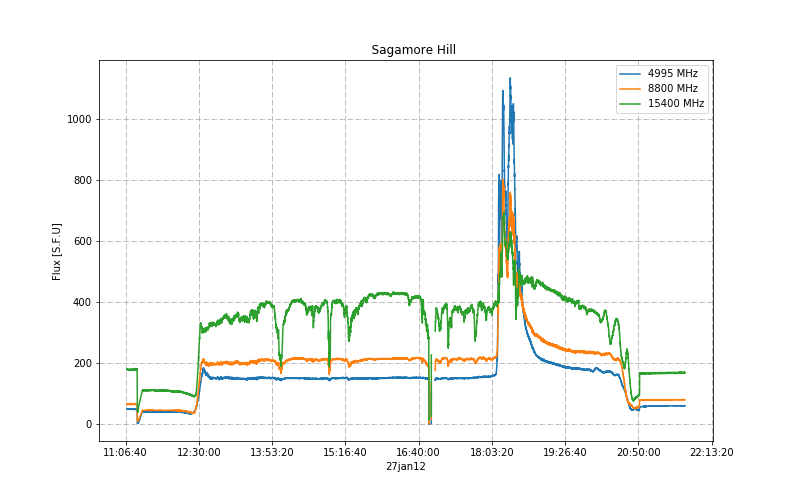}
\caption[RSTN data - Sagamore Hill and Palehua]{RSTN data for 2012/01/27 - Palehua (upper) and Sagamore Hill
(bottom)}
\label{fig:Fig_Cap6_RSTN} 
\end{figure}

To verify the existence of a solar event, we used data from the Radio Solar Telescope Network (RSTN) operated by the Meteorological Agency of the United States Air Force \citep{guidice1979}. This network is composed of 4 radio stations located around the globe. Considering the location of POEMAS in Argentina, we will use data from observatories in Palehua, Hawaii (USA), Sagamore Hill in Massachusetts (USA), and San Vito (Italy). The 3 observatories cover the POEMAS observation window depending on the time of year. However, the antenna with the highest time intersection is in Sagamore Hill, Massachusetts (USA). Data from Sagamore Hill and Palehua stations for 01/27/2012 are shown in Figure~\ref{fig:Fig_Cap6_RSTN}, in the bottom and top panels, respectively.

The solar flare that happened on this day at 18:15 UT is clearly seen in the RSTN data shown in the two panels of Figure~\ref{fig:Fig_Cap6_RSTN}. The impulsive phase of this event was also detected in the POEMAS signal at 45 GHz (Figure~\ref{fig:Fig_Cap5_Analise_ST}). However, to see the gradual phase of the flare, it is necessary to subtract the daily instrumental variation from the signal due to the telescope's misalignment. This can be done by subtracting the signal observed on the day before (or after) the event, since the variation in the signal does not vary significantly in the period of a day or so. However, this is not the procedure performed in this work due to its requirement of human supervision.

\section{Neural Network}\label{sec:NN}

Based on biological neural networks, that is, on the biological neuron, Artificial Neural Networks (ANNs) are mathematical models that have computational capacity acquired through learning and generalization. This structure attempts to mimic a human brain with connections between neurons (synapses) and input and output signals.

Frank Rosenblatt at the \textit{Cornell Aeronautical Laboratory} developed the first multi-neuron network of the linear discriminator type and named this network the \textit{perceptron}. A \textit{perceptron} is a network with neurons arranged in layers. This proposed model learns concepts and can answer with true (1) or false (0). In the early 1960s, Rosenblatt extended his work by publishing several articles and a book \citep{rosenblatt1962, Tappert2019}.

The resulting 45 GHz emission signal, after subtraction of the trend of the time series (black curve of Figure~\ref{fig:Fig_Cap5_Analise_ST}) was input to a Multilayer Perceptron Neural Network (NN). The temporal resolution of the light curve is 1 second, which would generate a lot of data points for the network. Therefore,  the temporal resolution was modified to 10 seconds, using the median in each 10-second interval to reduce the number of data points. Then a 5-minute window was considered to go through the resulting signal extracting chunks every 10 seconds. 

\begin{figure}
\centering
\includegraphics[width=.9\textwidth]{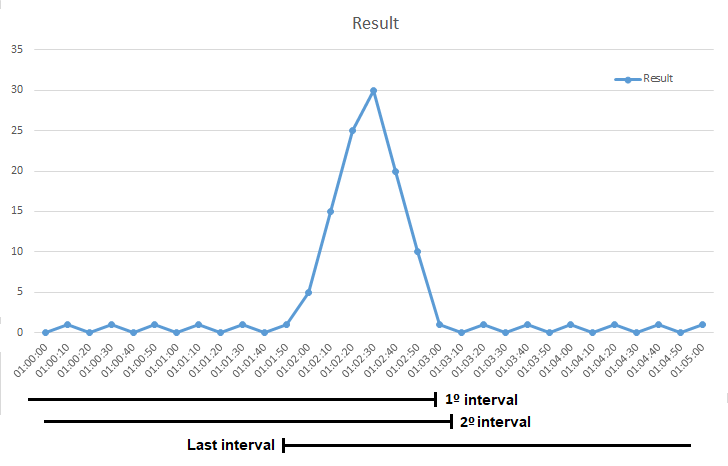}
\caption[Example of interval extraction]
{Example of interval extraction}
\label{fig:Fig_Cap6_Result} 
\end{figure}

To better exemplify the process, in Figure~\ref{fig:Fig_Cap6_Result} we show the validation of time intervals to be later considered as true signals for input to the neural network. In the example, the supposed event starts at 1:02:00 UT and ends at 1:03:00 UT. The first true interval would be [0:58:00, 1:03:00] UT, the second true interval would be [0:58:10, 1:03:10] UT, and so forth. Therefore, we would have a sliding 5 min window traversing the final signal every 10s. The last true interval of this supposed event would be [1:02:00, 1:07:00] UT. If this was the only event in one day, there would be 25 true intervals.

This was the most critical process in evaluating the signal due to the large volume of data. There were a total of 690 days of solar observation by the POEMAS with an average of 10 hours per day. Considering a temporal resolution of 10 s, there are approximately 3600 points in a day; thus resulting in $2.484 \times 10^6$ data points.

\subsection{Neural Network Training (NNT)}
Previously, \citet{ray2019Art} visually detected and analyzed 30 events observed by POEMAS telescope, which are listed in Table~\ref{tab:Tab_Cap7_Eventos}. First, we separate these events into two classes: training and classification. The first 14 flares were separated into the classification group. Events 15 through 30 were used for training of the neural network. The model adopted was supervised learning, where we submitted the 228 intervals referring to the 16 events and classified them as positive.

\begin{table}
\centering
\begin{tabular}{|c|c|c|c|}
\hline                   
N & Date & Time (UT) & NN \\
\hline                   
1 & 2011/12/25 & 18:14 & \checkmark \\
2 & 2011/12/25 & 20:27 & \checkmark \\
3 & 2011/12/26 & 20:16 & \checkmark \\
4 & 2012/01/27 & 18:15 & \checkmark \\
5 & 2012/03/05 & 19:13 & \checkmark \\
5 & 2012/03/06 & 21:05 & -- \\
7 & 2012/03/13 & 17:24 & \checkmark \\
8 & 2012/05/07 & 17:23 & --\\
9 & 2012/05/09 & 12:30 & \checkmark \\
10 & 2012/06/03 & 17:53 & \checkmark \\
11 & 2012/07/03 & 17:02 & \checkmark \\
12 & 2012/07/04 & 16:36 & \checkmark \\
13 & 2012/07/05 & 20:13 & \checkmark \\
14 & 2012/07/08 & 16:30 & \checkmark \\
15 & 2012/07/10 & 15:30 & --\\
16 & 2012/11/27 & 15:56 & \checkmark \\
17 & 2013/02/17 & 15:47 & \checkmark \\
18 & 2013/05/13 & 16:03 & \checkmark \\
19 & 2013/10/15 & 19:00 & -- \\
20 & 2013/10/25 & 15:00 & \checkmark \\
21 & 2013/10/26 & 19:25 & \checkmark \\
22 & 2013/10/28 & 15:10 & \checkmark \\
23 & 2013/10/29 & 21:48 & \checkmark \\
24 & 2013/11/05 & 18:10 & \checkmark \\
25 & 2013/11/05 & 22:11 & \checkmark \\
26 & 2013/11/06 & 13:43 & \checkmark \\
27 & 2013/11/07 & 12:26 & \checkmark \\
28 & 2013/11/07 & 14:18 & \checkmark \\
29 & 2013/11/07 & 14:23 & \checkmark \\
30 & 2013/11/07 & 14:29 & \checkmark \\
\hline                   
\end{tabular}
\caption{Events identified in the work of \citet{ray2019Art}. The last column lists the events identified in this work by the Neural Network (NN).}
\label{tab:Tab_Cap7_Eventos}
\end{table}

We need a balanced training base, and for that, we use the Nearmiss data balancing algorithm, a subsampling algorithm that randomly reduces most class examples. In the case of negative intervals, however, it selects samples based on distance \citep{mani2003}.

The next step is to define the NN structure after the training base is balanced (228 positives and 228 negative intervals). The input layer has 30 nodes that will receive each interval of 5 minutes with a resolution of 10 seconds. The output layer has only 1 node, which reveals if the  prediction for the data input is true or false. Several configuration tests were performed for the middle layer, using one and two layers. The best training results were found using only 1 intermediate layer. Thus, we decided to use one layer and varied the number of neurons to compare the results.

Because we are using supervised learning, when the NNT predicts a given input and the result is not expected, the network must adjust the weights to reduce the error and repeat the process until the error rate is 0. The activation function used was ReLu, as it is not linear and has better results compared to Tanh. 

ReLU is the most commonly used activation function when designing neural networks today.
The function is non-linear, and it does not activate all neurons at the same time, because if input is negative, it will be converted to zero and the neuron will not be activated. The Sigmoid function is a sensitive function and is continuously differentiable. This is not a linear function,  and this is an interesting feature because it essentially means that when
there are several neurons with sigmoid function as activation function the output too is nonlinear. This function varies between values [0,1]. The Tanh function is very similar to the sigmoid function. In fact, it's just an improved version of the sigmoid function as it varies between values [-1,1]
\citep{Burns2019}.

To assess the quality of training and classification, we consider accuracy as a figure of merit, as it defines the proximity of an experimental result to its actual value. The greater the accuracy, the closer it is to the actual result. For all NNT configurations performed, the training had an accuracy of 100\%. That is, they hit 228 positives (VP) and 228 negatives (VN) and had no false positives (FP) and no false negatives (FN).

\subsection{Classification}
The network efficiency and learning quality depend on its architecture specification, that is, the function of neuronal activation, learning rule, initial values, and training data. We consider a network with 3 layers: one input, one intermediate, and one output. This configuration showed the best performance and results in the training phase. For the primary classification, there were 14 events and 141 actual intervals. The 16 events used in the training were labeled in the classification as negative.

For each experiment, we varied the number of intermediate layer neurons. We started with 30 neurons and added 30 neurons in each experiment afterwards. In Table~\ref{tab:Tab_Cap5_RN}, we present the results of the 8 experiments. In the first column of the Table, the number of the experiment is given, in the second the number of neurons used in the intermediate layer, in the third column the number of positive intervals identified by the network as positive, and in the fourth column the number of positive intervals specified as negative. In the fifth column, the number of negative intervals identified as positive by the NNT are seen. Finally, in the sixth column, the number of negative intervals that are real negative, and in the seventh column, we have the accuracy of each of the experiments.

\begin{table}
\centering
\begin{tabular}{|c|c|c|c|c|c|c|}
\hline                   
& \# Neurons & True & False & False & True & \textbf{Acuracy}\\
& Occult layer & Positive & Negative & Positive & Negative & \\
\hline                   
\textbf{Exp. 1} & 30 & 126 & 15 & 1522161 & 747732 &  33\% \\
\textbf{Exp. 2} & 60 & 126 & 15 & 1334829 & 935064 &  41\% \\
\textbf{Exp. 3} & 90 & 127 & 14 & 1338661 & 931232 &  42\% \\
\textbf{Exp. 4} & 120 & 127 &  14 &  1246748 & 1023145 & 45\% \\
\textbf{Exp. 5} & 150 & 126 & 15 & 1522170 & 747732 &  33\% \\
\textbf{Exp. 6} & 160 & 126 &  15 & 1196951 & 1072942 &  47\% \\
\textbf{Exp. 7} & 170 & 131 & 10 & 1355206 & 914687 & 40\% \\
\textbf{Exp. 8} & 180 & 113 &  28 & 1764115 & 505778 & 22\% \\
\hline                   
\end{tabular}
\caption{Results of the Neural Network.}
\label{tab:Tab_Cap5_RN}
\end{table}

We started with 33\% accuracy in experiment 1 with 30 neurons in the middle layer. In experiments 2, 3, and 4 we had a gradual increase in accuracy, reaching 45\%. In experiment 5, with 150 neurons, we had a drop in accuracy to 33\%, which is equivalent to the results in experiment 1. As the results were not inferior to any previous investigation, we decided to continue increasing the neurons to check the results. In experiment 6 we achieved an accuracy of 47\% with 160 neurons in the hidden layer. In experiment 7, the accuracy decreases to 40\%, and experiment 8 reached the lowest accuracy found, 22\%. Experiment 7 had the highest number of true positives but a high number of false positives, so we focused on experiment 6 to analyze the results.
Experiment 6 with a configuration of 160 neurons had an accuracy of 47\%, the best among all the configurations. 

To improve the results, we note that each minute has 6 points (10 seconds time resolution) evaluated several times by the neural network. For experiment 9, if the network assessed less than 3 points within a minute as positive, we would consider that minute as negative. For experiment 10, if less than 4 points are evaluated as positive within one minute,  this minute is assumed to be negative. The proposed method improved the accuracy of the neural network to 60\%, as seen in Table~\ref{tab:Tab_Cap5_Metodo}. There are still many false positives and a considerable decrease in true positives, from 126 to 13, when comparing experiments 6 and 10. Also, the false negative increased from 15 to 128 cases.

\begin{table}[h]
\centering
\begin{tabular}{|c|c|c|c|c|c|c|}
\hline                   
& \# Positive  & True & False & False & True & \textbf{Acuracy}\\
& Points & Positive & Negative & Positive & Negative & \\
\hline                   
\textbf{Exp. 9} & $<3$ & 25 &  116 & 990033 & 1279860 &  56\% \\
\textbf{Exp. 10} & $<4$ & 13 & 128 & 906998 & 1362895 & 60\% \\
\hline                   
\end{tabular}
\caption{Results of the Neural Network for experiments 9 and 10.}
\label{tab:Tab_Cap5_Metodo}
\end{table}

Since the results of the neural network presented many false positives, we then compared the neural network results with the RSTN light curves, as well as the POEMAS observations, by visual inspection. First, the daily light curves from December 2011 to December 2013 (years of POEMAS observation) provided by RSTN were checked for events at frequencies $>4$ GHz (microwaves). When identifying a microwave event in the RSTN data, we check if the Neural Network identified this  event at the same day and time in the 45 GHz data.

\section{Results and discussion}\label{sec:results}
In this work, we analyzed the 45 GHz light curves observed by POEMAS telescopes for a total of 690 days, from December 2011 to December 2013. For the analysis, we considered the sum of the signal involving the two circular polarizations, RCP plus LCP. Moreover, visual inspection was performed on the daily microwave light curves observed by the RSTN telescope network during the same period.

The Neural Network (NN) application detected patterns in the 45 GHz light curve of the POEMAs and identified already known events and new ones. We compared the results of the Network with the work of \citet{ray2019Art}, who identified 30 events, as listed in Table~\ref{tab:Tab_Cap7_Eventos}. In addition, some events identified in the RSTN microwave data were not detected visually in the POEMAS light curves nor by the NN, given its temporal characteristic of long duration. Below we discuss each of these results in more detail.

\subsection{Events identified previously and confirmed by the Neural Network}
\begin{figure}[!h]
\centering
\includegraphics[width=.8\textwidth]{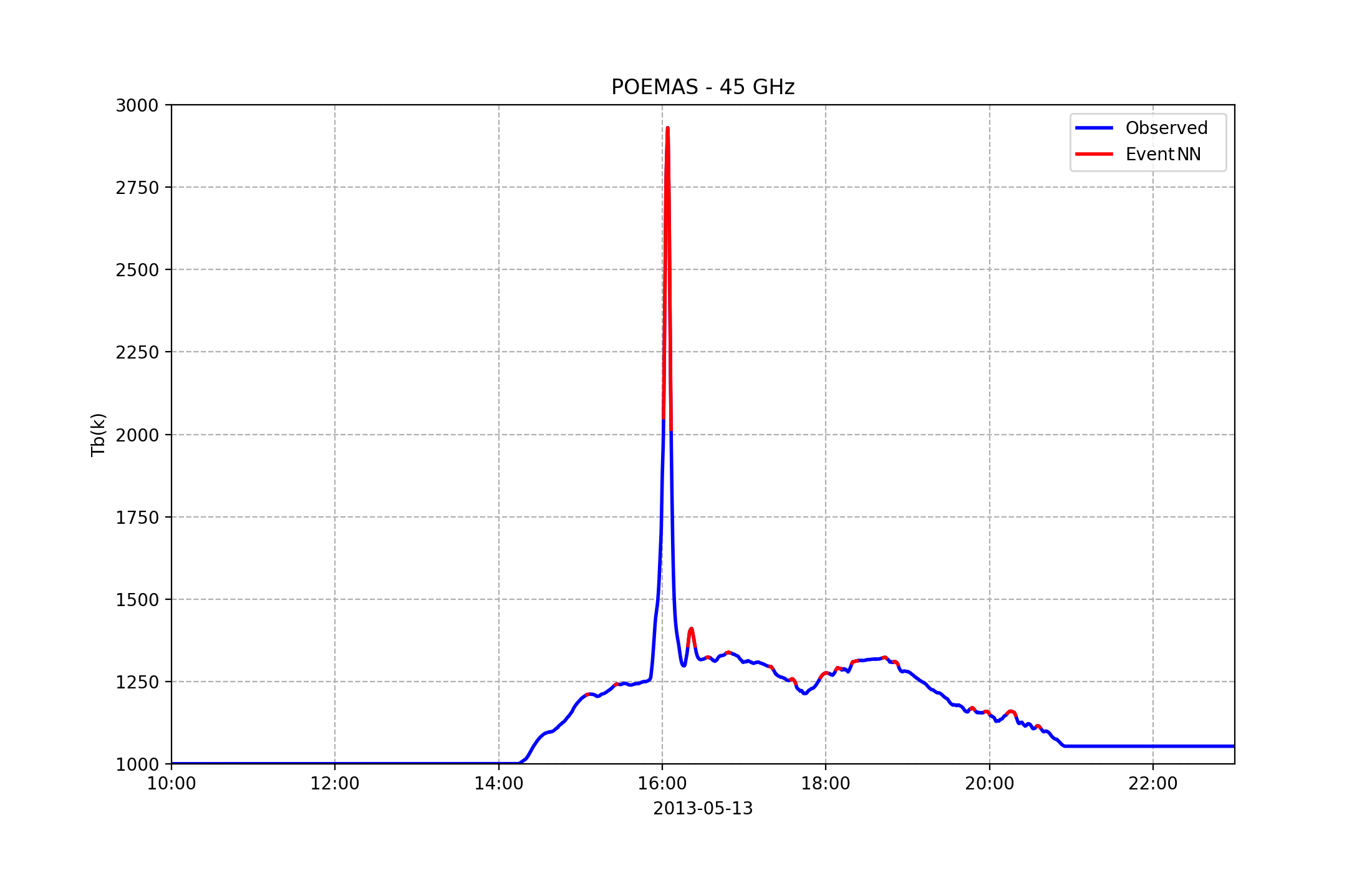}
\includegraphics[width=.8\textwidth]{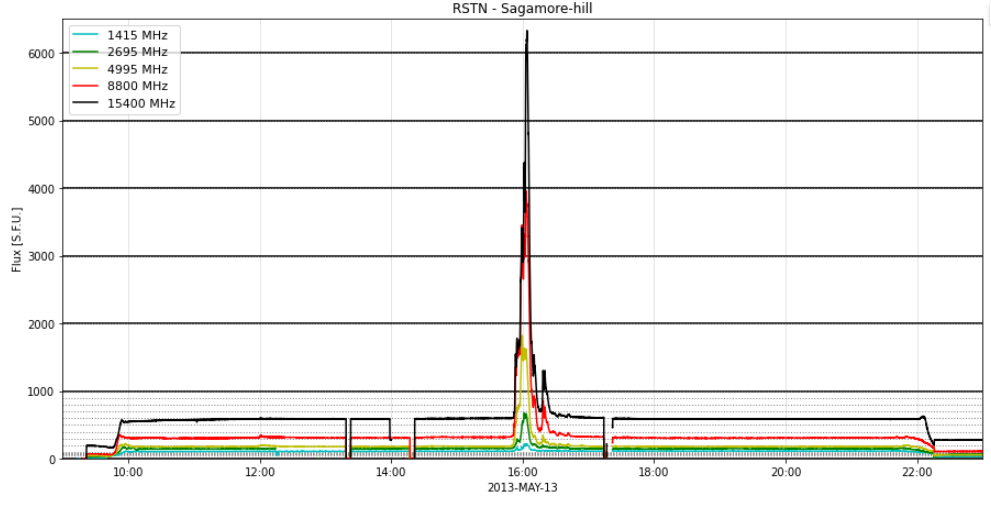}
\caption{Light curves from the event on 2013/05/13 at 16:03 UT. \textbf{Upper:} 45 GHz data from POEMAS (blue) and Neural Network result (red) and \textbf{Lower:} data from RSTN.}
\label{fig:Fig_Cap7_Eventos1}
\end{figure}

The Neural Network identified 26 of the 30 events from the work of \citet{ray2019Art}, thus the NN was able to retrieve 87\% of the events (last column of Table~\ref{tab:Tab_Cap7_Eventos}). 
An example of such an event is shown in Figure~\ref{fig:Fig_Cap7_Eventos1} that occurred on 2013 May 13th. In the top panel,  the POEMAS light curve is shown in blue, whereas the points identified by the NN as positive are depicted in red. In the bottom panel, the microwave light curves of the RSTN data from the Sagamore-Hill telescope (USA) are presented, where the event at 16:03 UT is clearly seen at all frequencies. Thus, the NN correctly identified the event that peaked at approximately 16:03 UT. This was a large event, of GOES X-ray class X2.8, that occurred on the East limb of the Sun.

\subsection{Events previously reported and not identified by the Neural Network}
\begin{figure}[!h]
\centering
\includegraphics[width=.8\textwidth]{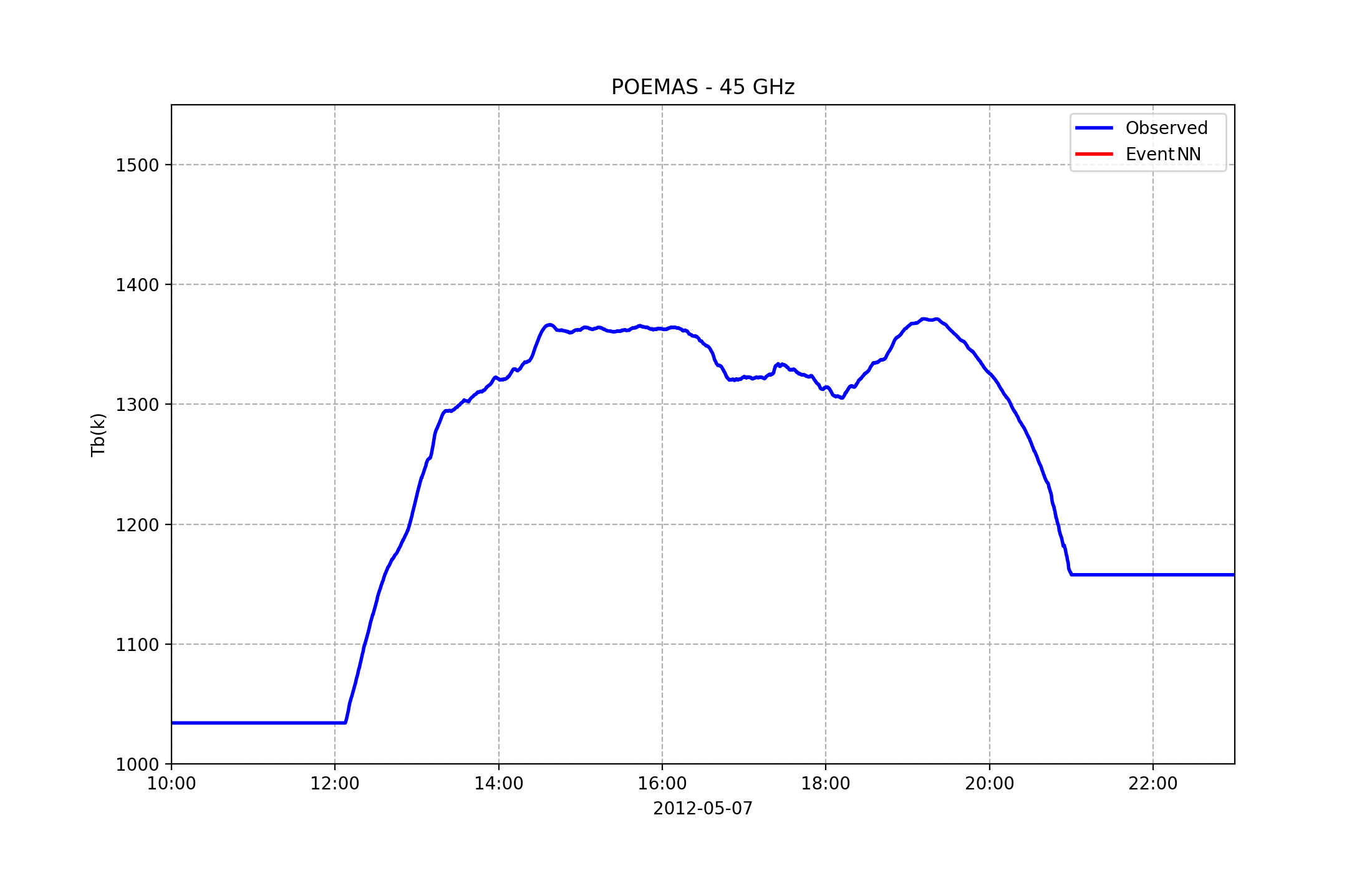}
\includegraphics[width=.8\textwidth]{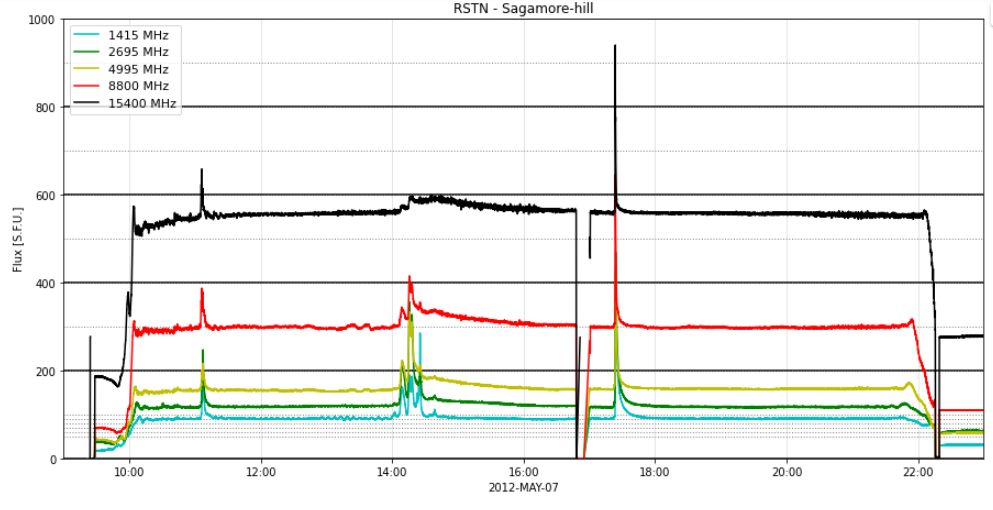}
\caption{Light curves from the event on 2012/05/07 at 17:23 UT. \textbf{Upper:} 45 GHz data from POEMAS (blue) and Neural Network result (red) and \textbf{Lower:} data from RSTN.}
\label{fig:Fig_Cap7_Eventos2}
\end{figure}

In Table~\ref{tab:Tab_Cap7_Eventos} there are four events identified visually by \citet{ray2019Art} but not recognized by the NN, which make up only 13\% of the events. The plots in Figure~\ref{fig:Fig_Cap7_Eventos2} represent the light curves of 2012 May 7th observed at 45 GHz by the POEMAS and in microwaves by RSTN. The event is clearly identified in the RSTN at approximately 17:23 UT. This event is not easily recognized in the POEMAS data and was neither identified by the NN, probably because of a duration longer than 5 min.

\subsection{New events identified by the Neural Network}
\begin{figure}
\centering
\includegraphics[width=.7\textwidth]{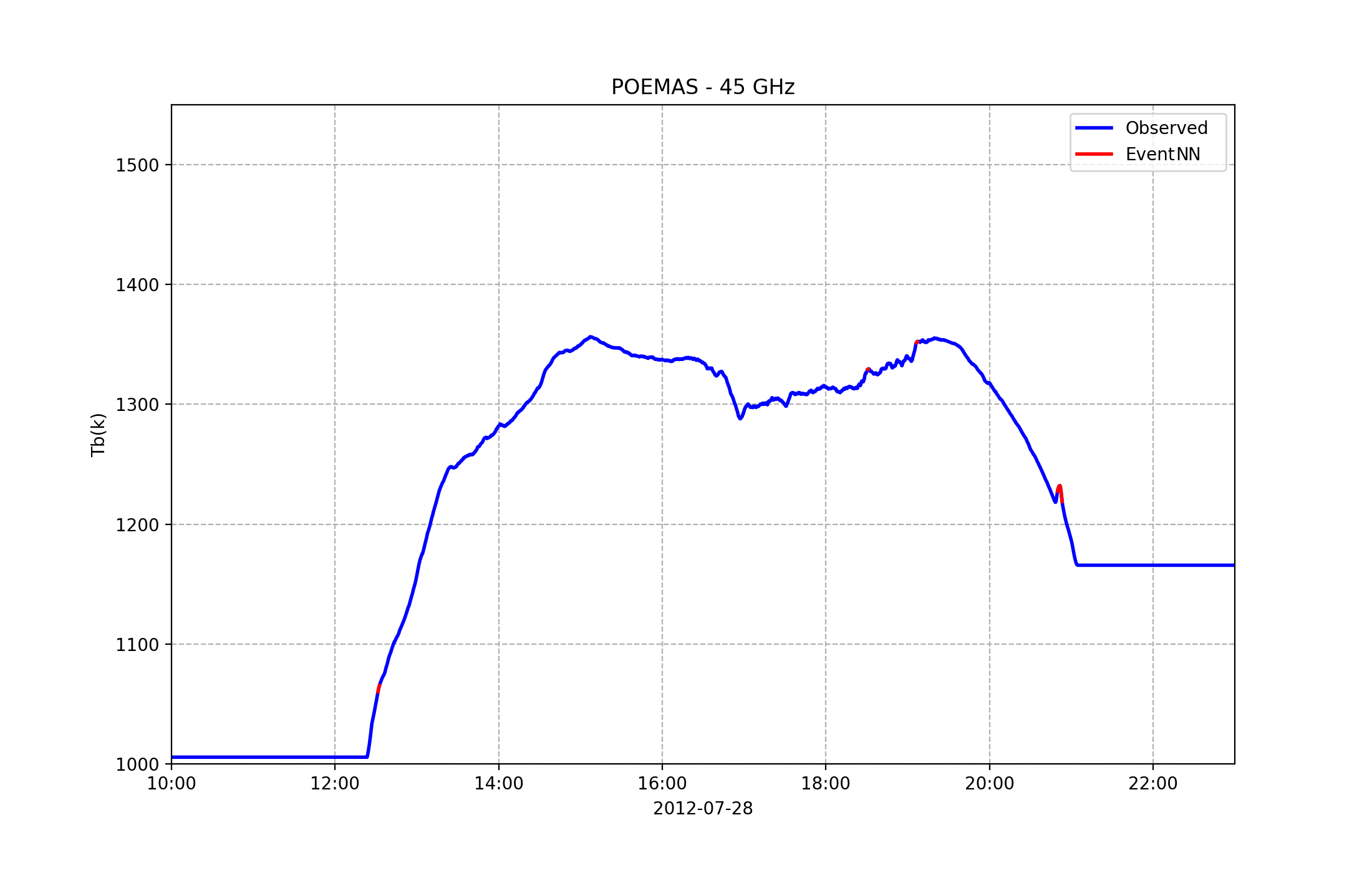}
\includegraphics[width=.7\textwidth]{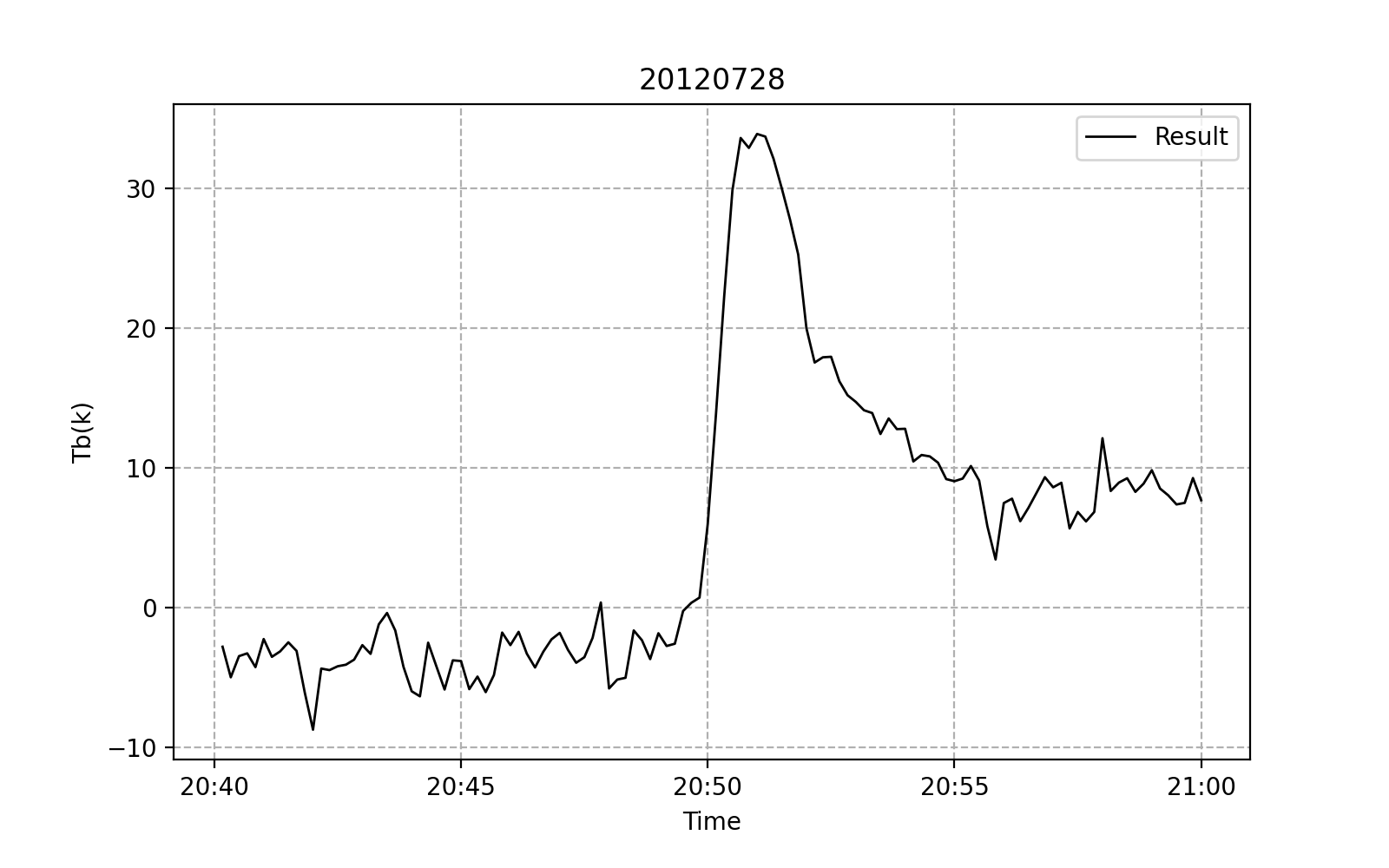}
\includegraphics[width=.6\textwidth]{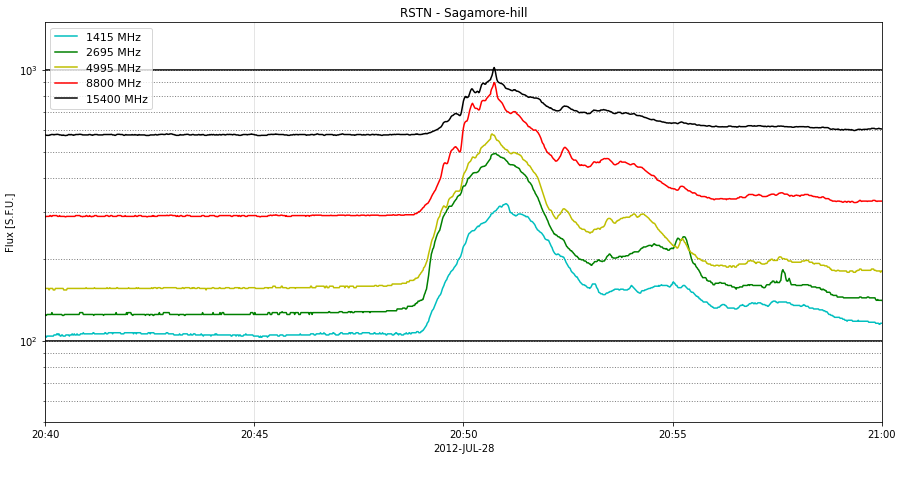}
\caption{\textbf{Top:} Poemas light curve on 2012 July 28th, with the event at 20:50 UT identified by the NN depicted in red. \textbf{Middle:} Highlight of the 45 GHz event after subtraction of the emission from the previous day. \textbf{Bottom:} Microwave light curve of the same day observed by RSTN.}
\label{fig:Fig_Cap7_RN}
\end{figure}

A total of 9 events were identified by the Neural Network, but went unnoticed in the visual inspection of \citet{ray2019Art}. These flares are listed in Table~\ref{tab:Tab_Cap7_Eventos3}. An example of such is the burst that occurred on 2012 July 28th,  identified by the NN at approximately 21:00 UT. The time profile of the flare is shown in the top right panel of Figure~\ref{fig:Fig_Cap7_RN}, where the light curve of the previous day was subtracted to eliminate the variation due to the misalignment of the telescope, and better show the event.

\begin{table}[h!]
\centering
\begin{tabular}{|c|c|c|}
\hline                   
N & Date & Time (UT) \\
\hline                   
1 & 2012/01/28 & 11:50 \\
2 & 2012/05/08 & 13:00 \\
3 & 2012/07/28 & 21:00 \\
4 & 2012/09/02 & 18:10 \\
5 & 2012/10/20 & 18:15 \\
6 & 2012/10/21 & 20:00 \\
7 & 2013/05/03 & 16:30 \\
8 & 2013/05/03 & 17:30 \\
9 & 2013/07/02 & 17:50 \\
\hline                   
\end{tabular}
\caption{Events identified only by the Neural Network}
\label{tab:Tab_Cap7_Eventos3}
\end{table}

\subsection{New long duration events visually identified}
\begin{figure}
\centering
\includegraphics[width=.7\textwidth]{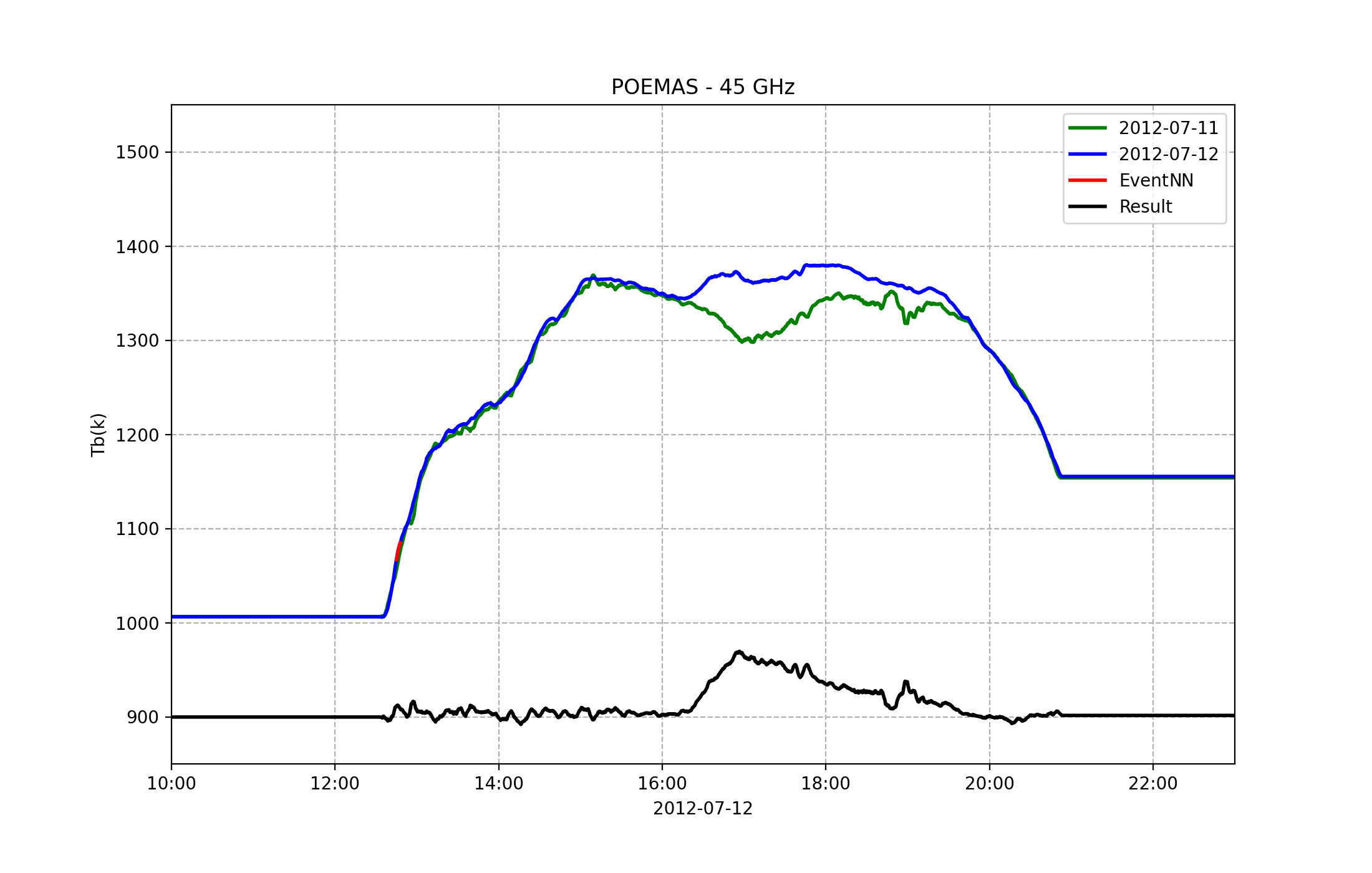}
\includegraphics[width=.7\textwidth]{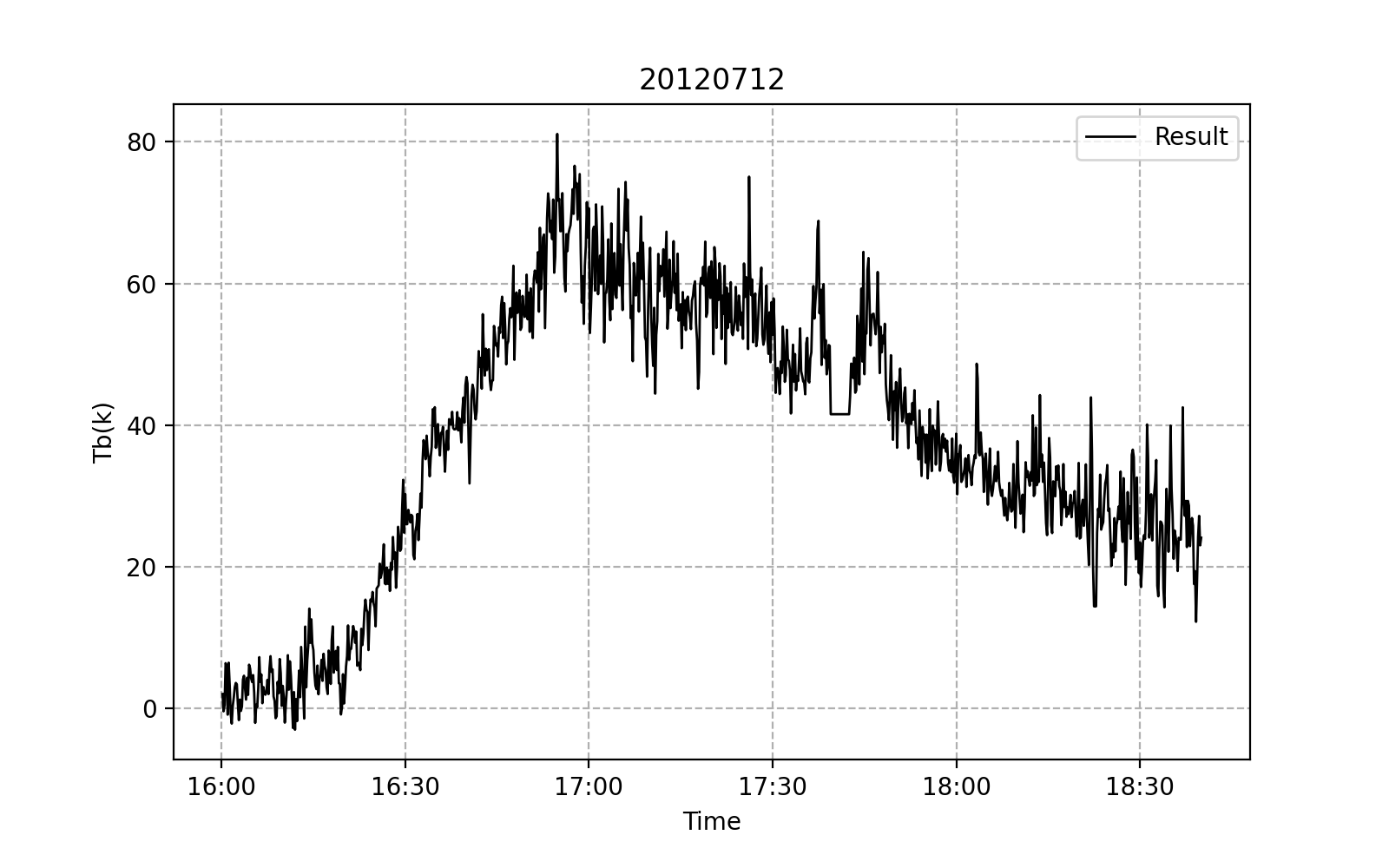}
\includegraphics[width=.6\textwidth]{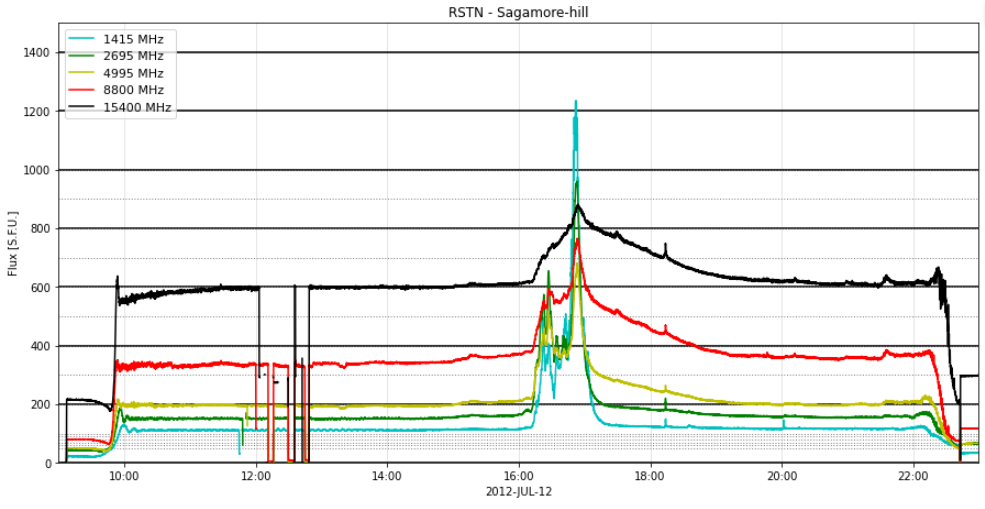}
\caption{Solar flare of 2012/07/12. \textbf{Top:} Light curve mission at 45 GHz from POEMAS for the day of the event, 2012/07/12 (blue curve) and the previous day, 2012/07/11 (green curve). The result of the subtraction of the emission on the 12th by the 11th of July 2012 is shown by the black curve (shifted by 900 to fit the scale of the figure). \textbf{Middle:} Blow up of the subtracted light curve to better identify the solar flare detected at 45 GHz. \textbf{Bottom:} Microwave light curves observed by  RSTN ($1 - 15$ GHz) for the whole day.}
\label{fig:Fig_Cap5_2012-07-12} 
\end{figure}

From the visual inspection of the RSTN microwave light curves, we identified 11 events not identified by \citet{ray2019Art} nor by the Neural Network, which are listed in Table~\ref{tab:Tab_Cap7_Eventos4}. An example of such event is shown in Figure~\ref{fig:Fig_Cap5_2012-07-12} for the 2012 July 12th flare, a long-duration event that lasted for more than 2 hours. The top panel shows the POEMAS 45 GHz light curve in blue, withe the light curve of the day before depicted in green. The subtraction of the emission from the previous day is shown in black on the top panel and highlighted in the middle panel of Figure~\ref{fig:Fig_Cap5_2012-07-12}. 
The flare that started at approximately 16:00 UT, peaked just before 17 UT and ended after 18:30 UT, was not readily visually identified in the POEMAS light curve.  In the bottom panel of the figure, the microwave light curves observed by RSTN clearly show the event at all frequencies, where the temporal profile of the 15 GHz emission closely resembles that of the 45 GHz from POEMAS.
The non-identification of this and the other 9 events probably occurred due to their gradual temporal profile, lasting from 30 minutes to more than an hour. We point out that the data input for NNT consisted of 5 min intervals.

\begin{table}
\centering

\begin{tabular}{|c|c|c|}
\hline                   
N & Date & Time (UT) \\
\hline                   
1 & 2012/03/02 & 17:40 \\
2 & 2012/03/03 & 18:00 \\
3 & 2012/03/04 & 11:00 \\
4 & 2012/03/17 & 20:50 \\
5 & 2012/06/06 & 20:00 \\
6 & 2012/06/14 & 13:30-15:00 \\
7 & 2012/07/12 & 16-17:00 \\
8 & 2012/07/27 & 17:15 \\
9 & 2013/07/03 & 20:00 \\
10 & 2013/08/17 & 18:20-19:30 \\
\hline                   
\end{tabular}
\caption{Long-term events not identified in the work of \citet{ray2019Art} nor by the Neural Network}
\label{tab:Tab_Cap7_Eventos4}
\end{table}

\section{Summary and conclusions}\label{sec:conclusion}
In this work, we analyzed two years of light curves from POEMAS telescopes at 45 GHz, from December 2011 through December 2013. The main objective was to automatically detect solar events in these light curves. The detection of solar flares in the light curves of POEMAS was hindered due to problems with the telescope's pointing, causing daily variations in the signal. Therefore, it was necessary to apply initial computational techniques to calibrate and reduce the POEMAS data. We created and used a Neural Network (NN) to identify solar flares in the data automatically. The application of this Neural Network was later compared with the microwave emission ($1 - 15$ GHz) detected by the RSTN radio-telescope network.

The first challenge of this work was the data transformation, due to noise and augmented by the misalignment of the telescope. Moreover, clouds obstructed the observation of the Sun and interference in the Earth's atmosphere even in the absence of clouds, such as increased water vapor or ice crystals, also precluded the detection of the flare signal by creating several spurious peaks in the light curves.

Using a Neural Network with supervised learning, we reached an accuracy of 47\%. This value is low, however it is due to the few samples used for training and the intrinsic noise of POEMAS' light curves. Later the accuracy was improved to 60\% by using a constraint on the False Positives. Nevertheless, despite the problems in the data mentioned above, we confirmed 26 previously known events and could identify 9 new events detected in the light curves of POEMAS, thanks to the NN.

Comparing the RSTN data for the two years of 2012 and 2013, we visually identified 10  long-term events not previously identified  in the POEMAS light curve by visual inspection nor by the NN. As the Neural Network was not supplied with any long-term events for training, it is not capable of  detecting this type of event. Thus the Neural Network constructed here can detect only short-term impulsive events, with duration less than 5 minutes.

\begin{table}
\centering
\begin{tabular}{|c|c|c|c|c|}
\hline                   
Event & Total & Old$^*$ & Neural Network & Table \\
\hline      
Old$^*$ & 30 & 30 (100\%) & 26 (87\%) & \ref{tab:Tab_Cap7_Eventos} \\
New & 9 & 0 (0\%) & 9 (100\%) & \ref{tab:Tab_Cap7_Eventos3} \\
Long duration & 10 & 0 (0\%) & 0 (0\%) & \ref{tab:Tab_Cap7_Eventos4} \\
\hline
Total & 49 & 30 (61\%) & 35 (71\%) & \\
\hline      
\end{tabular}
\caption{Summary of the events detected, or not, by the Neural Network.}
\small{$^*$ refers to the events identified visually by \citet{ray2019Art}.}
\label{tab:Resumo}
\end{table}

In summary, with the aid of artificial intelligence, in this work we have identified  35 solar events in the 45 GHz emission from 2012 to 2013 from a total of 49 bursts, or 71\%. If we consider, that the NN was not trained to detect events with duration longer than 5 min, then the accuracy of the NN increases to 90\%.
From the total of 49 events, 19 solar flares are unprecedented at 45 GHz, not having been identified in previous works that analyzed these data \citep{ray2019Art}. The statistics of the NN is summarized in Table~\ref{tab:Resumo}.

The use of artificial intelligence is innovating Astronomy. Especially in Solar Physics, we can mention the works by \citet{hou2020, ishikawa2021, neira2020},
 who obtained an accuracy of approximately 90\%. To leverage the study of solar flares, such techniques must be explored.

The search for solar flares using Neural Network was a first step to automating the process. Several challenges were encountered, such as clouds, periodic signal variations, and non-detection of long-term events, and their solution will be proposed in future work.



\section*{Acknowledgements}
A.V acknowledges partial financial support from  FAPESP grant \#2013/10559-5. V.L. thanks the fellowship from MackPesquisa funding agency.
\appendix



\bibliographystyle{elsarticle-harv} 
\bibliography{bibliografia}





\end{document}